\documentclass[journal]{IEEEtran}

\usepackage{graphicx}
\usepackage{multirow}
\usepackage{amssymb}
\usepackage[cmex10]{amsmath}
\usepackage{cite}
\usepackage{subfig}
\usepackage{gensymb}
\usepackage[T1]{fontenc}
\usepackage{mathptmx}
\usepackage[scaled]{helvet}    
\usepackage{luximono}          
\usepackage{txfonts}

\def\paren#1{\left({#1}\right)}
\def\brack#1{\left[{#1}\right]}

\def\set#1{\left\{{#1}\right\}}

\def\tn{{\tilde n}}

\def\tt{\tilde t}

\def\set#1{\left\{{#1}\right\}}
\def\paren#1{\left({#1}\right)}
\def\brack#1{\left[{#1}\right]}

\def\prob#1{P\left\{{#1}\right\}}

\newcount\exnum

\long\def\bexer{%
\global\advance\exnum by1
\global\insertnum=\exnum
\binsert{Exercise}}

\begin{document}
%

\title{A molecular communications model for drug delivery}

\author{Mauro~Femminella, Gianluca~Reali, Athanasios V. Vasilakos$^*$%
\thanks{This work was supported in part by EU project H2020 FET Open CIRCLE (Coordinating European Research on Molecular Communications). Asterisk indicates corresponding author.}
\thanks{M. Femminella, and G. Reali are with the Department of Engineering, University of Perugia, via G. Duranti 93, 06125 Perugia, Italy, (email: \{mauro.femminella,gianluca.reali\}@unipg.it).} 
\thanks{A. V. Vasilakos is with Department of Computer Science, Electrical and Space Engineering, Lulea University of Technology, Sweden (email: vasilako@ath.forthnet.gr).}}  

\maketitle
\begin{abstract}
This paper considers the scenario of a targeted drug delivery system, which consists of deploying a number of biological nanomachines close to a biological target (e.g. a tumor), able to deliver drug molecules in the diseased area. Suitably located transmitters are designed to release a continuous flow of drug molecules in the surrounding environment, where they diffuse and reach the target. These molecules are received when they chemically react with compliant receptors deployed on the receiver surface. In these conditions, if the release rate is relatively high and the drug absorption time is significant, congestion  may happen, essentially at the receiver site. This phenomenon limits the drug absorption rate and makes the signal transmission ineffective, with an undesired diffusion of drug molecules elsewhere in the body. The original contribution of this paper consists of a theoretical analysis of the causes of congestion in diffusion-based molecular communications. For this purpose, it is proposed a reception model consisting of a set of pure loss queuing systems. The proposed model exhibits an excellent agreement with the results of a simulation campaign made by using the Biological and Nano-Scale communication simulator version 2 (BiNS2), a well-known simulator for molecular communications, whose reliability has been assessed through in-vitro experiments. The obtained results can be used in rate control algorithms to optimally determine the optimal release rate of molecules in drug delivery applications.
\end{abstract}
%
%


\section{Introduction}\label{intro}

Molecular communication is a novel paradigm allowing information exchange between biological nanomachines (bio-nanomachines) over short ranges, within an aqueous environment \cite{Akyildiz08}. In this model, a transmitter nanomachine emits molecules, which propagate and eventually are received by a receiver nanomachine. The information is usually encoded in either the timing or the amplitude of the concentration of molecules. The reception process usually consists of a chemical reaction between those molecules (ligand) and compliant receptors present on the receiver surface. Bio-nanomachines are made of biological materials and can perform a number of simple tasks which need to be coordinated, by means of message exchange, in order to accomplish complex functions.

Diffusion-based molecular communications are a special form of molecular communications \cite{Nakano12c}, very common in nature. They can be roughly classified in two categories, namely pure-diffusion and diffusion with drift. For instance, the former includes communications in the extracellular matrix of the connective tissue, whilst the latter encompasses communications in circulatory and lymphatic systems.

In this paper, we consider drug delivery in pure-diffusion molecular communications. In drug delivery systems a nearly continuous flow of molecules is transmitted from the transmitter (TX) to the receiver (RX) \cite{Allen04}, in order to achieve a desired effective average drug delivery rate at a specific target. In these applications, the TX node has the role of an actuator, and the RX node is either the target or a sensor which measures and controls the delivery of drug/grow factors molecules to the surrounding cells.

In this scenario, the space around the receiver could become full of drug molecules, which produce congestion. Congestion control in packet networks is a research topic that attracted researchers in the last 30 years (see, e.g., \cite{Winstein:2013:TEM:2486001.2486020,Sivaraman:2014:ESL:2619239.2626324}), since the seminal works of Van Jacobson \cite{Jacobson:1988:CAC:52324.52356} and K.K. Ramakrishnan \cite{Ramakrishnan:1990:BFS:78952.78955}. However, until now, little attention has been paid to the issue of congestion in molecular communications, beyond some initial investigations reported in \cite{nakano13,TCPm14}. Indeed, molecular communications, especially those making use of diffusion-based propagation \cite{Philibert2006}, are not only very sensitive to congestion, but also difficult to control due to the high communication latency. Thus, an early detection of congestion is necessary in order to avoid heavy receiver overload. However, beforehand, we have to give a definition of congestion in molecular communications.

In molecular communications, congestion consists in the inability of receivers to capture all molecules in the receptor space. This limitation is due to the combination of two factors, which are the number of receptors present over the receiver surface and the trafficking time \cite{Lauffenburger1993}, which can be regarded as the molecule reception time. The trafficking time is defined as the time needed to make a binding between a ligand and a compliant receptor, plus the time to internalize the resulting complex, plus the time to re-expose another receptor. It can last up to tens of seconds \cite{Lauffenburger1993} and significantly affect the molecule absorption process. 

\textbf{Motivation}: in the bio-medical field, the receiver congestion problem is referred to as ``receptor saturation''. Building a model for receptor saturation is a very important open issue, since the lack of a general model requires executing expensive lab experiments for receptors characterization \cite{pmid21760560}. In addition, receptors saturation affects drug delivery and disease modeling \cite{pmid22830443}. However, most of research in the molecular communications 
either neglect the presence of individual receptors (absorbing receiver \cite{6807659}), or neglect the absorption time of molecules (absorbing receptors \cite{6967753}). A more accurate model, which partially addresses these shortcomings by modeling molecules absorption via receptors with a birth-death Markov process, is presented in \cite{Pierobon2011sept}. However, also this model has its shortcomings. In fact, it models receptors as a pool of servers, which can be invoked and used when any new molecule enters the system. Instead, receptors are isolated service elements, and this nature must be taken into account.

\textbf{Main contributions}: the paper has two main contributions. The first one is the analysis of the causes of congestion at RX when a continuous flow of molecules is transmitted by the TX. The second and main contribution is a queuing model that includes these causes, which has been validated by an extensive simulation campaign. Our idea is to model each receptor on the RX surface as a pure loss queuing system. Specifically, we model each receptor by an M/M/1/1 queue, where the mean service time is the mean trafficking time. Given the unguided diffusion of the transmitted signal, 
we model the receiving nanomachine as a node with a large number of receiving antennas (receptors), each interfaced to a layer 2 with very high processing time (the trafficking time) without waiting room. Thus, when the concentration of molecules nearby the RX surface becomes excessive (channel congestion) and a significant part of receptors are busy, also the RX becomes congested (receiver congestion). This analogy between diffusion-based molecular communications and  wireless communications is motivated also by recent papers dealing with multiple-input and multiple-output schemes \cite{mimo} and directional antennas \cite{antenna} to build directional receivers in the field of molecular communications, similarly as in the wireless world. 

Finally, although the queuing theory has already been used in the field of bio-physics for modeling ligand-receptor systems \cite{McQuarrie,Lauffenburger1993}, the novel contribution of this paper consists of using this theory for modeling receptor saturation from the communications perspective, thus treating it as a congestion problem in communications networks. To the best of our knowledge, this is the first paper using this approach.

The paper is organized as follows. In section \ref{back}, we illustrate the related work in the field. Section \ref{model} presents the system model, including a detailed description of our proposed model, which is able to capture the root causes of congestion. Numerical results, which include the results of the simulation campaign, used to validate the proposed mechanisms, are presented in Section \ref{perf}. Finally, we draw our conclusions in Section \ref{conclu}.

\section{Background and Related Works}\label{back}
\subsection{Molecular communications}
A complete view of a layered network architecture for molecular communications has been proposed in \cite{Vasilakos14}. Following the layered architecture of traditional communication networks, such as the Open Systems Interconnection model (OSI) and TCP/IP reference model, a formal model was developed for each layer, and potential research directions are illustrated for each of them. Other works explore issues related to the high protocol layers, including connection-oriented protocols\cite{TCPm14}, feedback-based rate control \cite{nakano13}, network coding \cite{6867315}, and security issues \cite{6879470}.

However, major research efforts in molecular communications are focused on the physical layer issues of different communication media. In particular, the information capacity and the physical features (e.g., delay, signal attenuation and amplification) of molecular communications are studied by using random walk models \cite{Moore09,Nakano12d,Kuran201086}, random walk models with drift \cite{Kadloor2012,Srinivas12}, diffusion-based models \cite{Akan08,Atakan10,Mahfuz10,Pierobon10,Pierobon11c}, diffusion-reaction-based models \cite{Nakano10a,Nakano11b}, active transport models \cite{Moore09,6766704}, and a collision-based model \cite{Guney2012}. 

As for the diffusion-based models, which is the model considered in this work, a review of different transmission schemes is provided in \cite{ShahMohammadian2012}, where the existing schemes are classified into pulse position modulation (PPM, \cite{Kadloor2012}), concentration shift keying (CSK, \cite{Atakan2010}), and  molecule shift key (MoSK), a further proposed category which combines bursts of different molecules to encode data. The possible trade-off between symbol duration and communication distance is analyzed in \cite{6612809}. 

Some recent papers show enhanced receivers for diffusion-based molecular communications \cite{lltatser13, Kuran201086,6612809,6906290,6881686,6782407,6708551,Yilmaz2014136}. However, these works, as most of research papers in molecular communications, do not take into account that signal reception occurs via binding between a signal molecule and the relevant surface receptor, and the number of receptors is typically assumed to be so large to cover all the receiving nanomachine surface. This is referred as the so-called \textit{absorbing} receiver model \cite{6807659}, in which a molecule, when gets in touch with the surface of the RX node, is immediately absorbed and removed from the surrounding environment. Although the absorbing receiver model is acceptable for some processes used by cells to engulf other cells, microvesicles, or proteins (phagocytosis and/or pynocytosis, \cite{pmid23584393}), most of cells communications are carried out via receptors. Thus, in some situations the assumption of using an absorbing receiver is a very rough approximation, since receptors are a finite resource and each nanomachine usually has a limited, yet large, number of receptors of a given type. A variation of the absorbing receiver model is the receiver with absorbing receptors \cite{6967753}, in which a molecule is instantaneously absorbed when it hits a compliant receptors. Again, although this model is acceptable in some specific situations (e.g. for modeling ions channels \cite{pmid19339978}), in most of cases it does not, since the stochastic nature of the ligand-receptor binding is not considered. Differently, it is contemplated in \cite{Pierobon2011sept}, and further investigated in \cite{daigle14}, which also includes the trafficking time \cite{Lauffenburger1993} in the reception process. 

\subsection{Drug delivery systems}
One of the most popular topics in nanomedicine is the targeted drug delivery, since it could be the basis for the modern medical therapeutics. Its main goal is to provide a localized drug delivery only where medication is needed, thus avoiding to affect other healthy parts of the body. This is due to the very small size of the particles delivering medicine, which is in the order of nanometers (nanoparticles), which allow them to diffuse into the bloodstream and across the vascular and interstitial barriers. Nanoparticles present ligands on their surface, in order to improve cell targeting. In fact, these ligands increase the chances of binding to the surface receptors of target cells \cite{Elias2013194}.

Many types of drug delivery, based on molecular communications, have been proposed in the technical literature, with both passive and active transport of molecules. For what concerns passive transport, the most popular approach is based on particle diffusion, with or without drift. In case the drug delivery happens via the circulatory or lymphatic system, the most appropriate model is the diffusion with drift, by also considering the effect of collisions with blood cells \cite{Tan2012,Felicetti201398}. In case the drug delivery happens outside blood vessels, e.g. in the extracellular matrix, the mere diffusion-based model is sufficiently accurate. In case of active transport, the main alternatives are those based on bacteria.

In order to design an efficient and targeted drug delivery, it is essential to correctly determine the distribution of particles over time. For this reason, the relevant research has produced several approaches, from statistical models \cite{PhysRevE.86.021905} to analytical ones \cite{6548006}. In particular, drugs delivered through bloodstream can reach any part of the body, by passing through the complex network of blood vessels (Particulate Drug Delivery System, PDDS). In \cite{6548006}, further expanded in \cite{7103028,7031911}, the authors propose a specific model of the cardiovascular system, by analyzing the peculiarities of different vessels, from the smallest to the largest ones, thus obtaining a drug propagation network model which takes into account even bifurcations and junctions of vessels. The resulting model includes also the individual features  of the cardiovascular system by physiological parameters, such as the heartbeat rate profile. This model could be useful to guarantee that the drug concentration does not reach toxic levels in the body and it is essential for optimizing the drug injection, by identifying the suitable injection points that guarantee the best delivery profile for maximizing the treatment benefits while minimizing the amount of drugs in the other parts of the body. To this aim, it is necessary to realize a complete control of the drug release, from the initial release, to the suitable release rate. 

A different and even more innovative model considers the delivery from a number of nanomachines (nano-actuator), implanted close to the target (e.g. a tumor), which bypasses the injection through the cardiovascular system, thus minimizing the side effects on the healthy parts of the body. A theoretical model on the local rate control between a nano-controller and a nano-actuator has been presented in \cite{nakano13}. However, this is a high abstraction model, which does not consider fine grained dynamics and does not explore the receiver congestion phenomenon. Subsequently, a related communication protocol, inspired by the TCP congestion control, has been proposed in \cite{TCPm14}. This protocol can control the drug release process by using feedback messages.  A challenging aspect of the protocol operation is the usage of feedback messages, which makes the protocol mostly suitable for environments without drift. 


Drug delivery systems lay their foundations on the interactions of drug molecules with the receptors at target cells \cite{Colquhoun2006149,Kenakin2004186,pmid16402126}. Two opposite main theories exist, namely the occupancy theory \cite{clark1933} and the rate theory \cite{paton1961}. The former assumes that the drug effects are maximized when all receptors at the target are occupied by the drug, whereas the rate theory is developed on the assumption that the excitation of a stimulant drug is proportional to the rate of the drug-receptor combination, rather than to the proportion of receptors occupied by the drug. However, behavior of drugs can be quite different. For instance, it is not always necessary to reach an occupancy of 100\% of receptors to produce a full response on the target cell \cite{Lambert01122004}. In addition, it was found that a good model for describing the effects of the drug-receptor interaction depends on the type of drug \cite{O'Brien1987327}. In any case, our proposed receiver model can be used for analyzing both of them.

\section{The System Model}\label{model}

\begin{figure}[!t]
\centering
\includegraphics[width=0.90\linewidth]{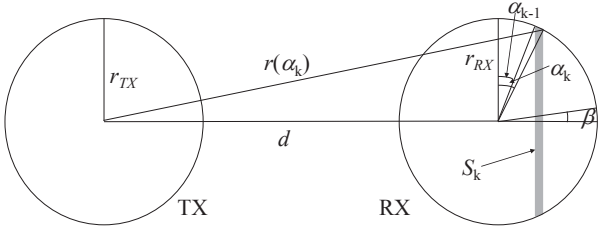}
 \caption{System model. We assume there is a circular symmetry around the axis $d$.}
 \label{layout}
\end{figure}

The considered drug delivery scenario consists in two fixed bio-nanomachines, staying at a distance $d$ between their centers, as depicted in Fig. \ref{layout}. One of them is the transmitting node and the other is the receiver. The communication happens by means of signal molecules released by TX, which propagate by diffusion, modeled as Brownian motion \cite{Philibert2006}.  The Brownian motion is characterized by the diffusion coefficient, given by $D=\frac{K_b T}{6 \pi \eta r_{c,tx}}$. $K_b$ is the Boltzmann constant, $T$ is the temperature expressed in Kelvins, $\eta$ is the viscosity of the medium, and $r_{c,tx}$ is the radius of the considered molecules. The shape of both RX and TX is spherical with radius $r_{RX}$ and $r_{TX}$, respectively. The number of receptors deployed on the RX surface is $R_{RX}$, and they are modeled as a circular area of radius $r_{r,rx}$. 

The transmission is organized in bursts of molecules of size $Q$, spaced in time by a short period equal to $\Delta t$, so as to emulate a continuous emission of drug molecules with rate $Q/\Delta t$. The reception process at receptors is modeled as an exponential random variable with mean $T_{traff}$. We assume that the communication session can be set up and torn down with a specific protocol, e.g. similar to the one shown in \cite{TCPm14}. However, since our results are unaffected by the protocol used, the latter is beyond the scope of this paper.

Let us assume that the center of TX is located at the origin of the system of coordinates, and initially we neglect the disturbing presence of the RX. We recall that the RX is not absorbing, although able to capture molecules. We now solve the diffusion equation upon the transmission an impulse of size $Q$ molecules at $t=0$. The concentration of molecules $c(t,r)$, in a given point of the space at a distance $r$ from the center of system of coordinates and at a specific time $t$, can be obtained by solving the diffusion equation (Fick's second law of diffusion \cite{Philibert2006}), which is 
\begin{eqnarray}\label{fick_eq}
\frac{\partial c(t,r)}{\partial t}  &=& D \nabla^{2} c(t,r),
\end{eqnarray}
\noindent obtaining the system impulse response scaled by $Q$:
\begin{equation}
	h_Q(t,r)= \frac{Q}{\left(4 \pi D t\right)^\frac{3}{2}} e^{\left(-\frac{r^2}{4D t}\right)}.          
\label{conca}
\end{equation}

Since the diffusion equation is a linear equation \cite{6648629}, the solution to (\ref{fick_eq}) for a train of bursts of size $Q$ spaced by a period $\Delta t$ ($\sum_{i=1}^{\left\lfloor t/ {\Delta t} \right\rfloor} {Q \delta(t - i \Delta t)}$) and started at $t=0$ is 
\begin{equation}
c(t,r)=\sum_{i=1}^{\left\lfloor t/ {\Delta t} \right\rfloor} { h_Q(t - i \Delta t,r) } \approx \frac{1}{\Delta t} \int_{0}^{t} {h_Q(\tau,r) d\tau}\xrightarrow[t \rightarrow \infty]{}\frac{Q}{{\Delta t} 4 \pi D r}.          
\label{train}
\end{equation}
This is a significant result, although obtained with the simplifying assumption of neglecting the presence of the RX node. The arrival rate of molecules in the receiver space depends on the concentration close to the receiver surface \cite{Pierobon2011sept}, which, in turn, from (\ref{train}), results to be inversely proportional to the distance from the source of molecules, that is $c(r) \propto \frac{1}{r}$.


As mentioned above, most of the receiver models do not take into account the presence of receptors on the receiver surface, and assume that the receiver can capture all compliant molecules in the surrounding space (absorbing receiver). Clearly, these models are unsuitable for detecting congestion since 
it is due to the limited rate of creation of ligand-receptors bonds, and depends on both the number of receptors $R_{RX}$ and on the trafficking time $T_{traff}$. 

In this regard, a more accurate model is the ``reversible first-order reaction'' model presented in \cite{Pierobon2011sept}. Although this model has been proposed for a different goal, it is instructive to review it in the perspective of congestion in molecular communications. In addition, it allows further detailing the system model.
\subsection{Review of reversible first-order reaction model}\label{review}
According to \cite{Pierobon2011sept}, the status of the receiver at any time $t$ can be represented by the value of a random variable $\tn(t)$, which denotes the number of ligand-receptor bonds at time $t$. Thus, the receiver behavior is modeled by the random process $\set{\brack{\tn(t)}, t \geq 0}$. In general, $\tn(t)$ increases when new  ligand-receptor bonds are created, and decreases due to the internalization of these bonds \cite{Lauffenburger1993}, that is ligand molecules assimilation. In our model, we neglect the probability of the secondary effect of the rupture of the ligand-receptor bond before its internalization, in order to simplify the comprehension of the main underlying phenomena. However, if needed, it can be easily introduced in the receiver model.

The random process $\set{\brack{\tn(t)}, t \geq 0}$ is modeled as a birth-death, discrete-valued, continuous-time Markov process, with $\tn(t) \in \set{0, 1, 2, \dots, R_{RX}}$ \cite{Kleinrock:1975:TVQ:1096491}.  Let $\pi_{i}(t)$ denote the probability that the Markov chain is in state $i$ at time $t$; that is,  $\pi_{i}(t) = \prob{\tn(t)=i}$. This means that the system has $i$ active ligand-receptor bonds. Then the dynamical equations for the state probabilities are as follows:
\begin{eqnarray}\label{GrindEQ__5A_}
\frac{d}{dt} \pi_{i}(t) &=& \lambda_{i-1}\pi_{i-1}(t)-\paren{\lambda_i+ \mu_i} \pi_{i}(t)+ \mu_{i+1} \pi_{i+1}(t),
\end{eqnarray}
where is noted that all the rates $\lambda_i$ and $\mu_i$ are state-dependent. In particular, we observe that
\begin{itemize}
\item the coefficient $\lambda_i$ describes the bond formation rate when the number of active bonds are $i$. In a queuing system, it represents the users arrival rate when the system is in state $i$. It depends upon the number of available receptors on the cells surface, $R_{RX}-i$, the concentration of molecules close to the cells surface at a given time, $c(t,r)$, and on the per receptor association rate constant, ${k}_{+}$  \cite{Pierobon2011sept,Kuo1997}. Specifically, the constant  ${k}_{+}$ in \cite{Pierobon2011sept} is defined as the product of $Z$, which represents how frequently a collision between a particle and an unbound receptor happens, and $F_C(E_{a})$, which is the fraction of collisions having an energy higher than the threshold $E_a$, so that they can form a bond. Clearly, it results that $Z$ is proportional to the average number of free receptors, that is $Z \propto (R_{RX}-n_b)$, where $n_b$ is the average number of bond receptors in steady-state conditions. In turn, $n_b$ is a function of the average arrival rate, $\lambda_a$, formally defined in (\ref{arrival}). As for $F_C(E_{a})$, in our model, each collision between a ligand and a compliant receptor corresponds to an assimilation, thus  $F_C(E_{a})=1$. The resulting rate is the product of these three quantities, that is,
\begin{equation}\label{Lambda_def}\lambda_i = {k}_{+}\ c(t,r) \paren{R_{RX}-i} = Z c(t,r) \paren{R_{RX}-i};\end{equation}
\item the coefficient $\mu_i$ describes the bond internalization rate when the number of active bonds are $i$. A memoryless model for this phenomenon is widely accepted in literature \cite{Lauffenburger1993}. It depends upon the number of currently formed complexes and on the average trafficking time, where the trafficking time is a random variable exponentially distributed with average value $T_{traff}$. This means that  
\begin{equation}\label{mu_def}\mu_i =  \mu  i =  i / T_{traff}.\end{equation} 
\end{itemize}
We assume that the TX node sends molecules according to the constant pattern with rate $Q/ \Delta t$. Then, in the steady state, the arrival process at the receiver can be approximated as a Poisson process, due to the randomness and memoryless properties of Brownian motion. The average arrival rate $\lambda_a$ observed by the receiver is given by
\begin{equation}\label{arrival}\lambda_a = \sum_{i=0}^{R_{RX}-1} {\lambda_i \pi_{i}}.\end{equation}
However, in addition to the arrival rate, also the value of the rejection rate $\lambda_r$ is very important. In fact, since the model consists of a pure loss queuing system with $R_{RX}$ servers and state dependent arrival rate, it clearly exhibits some losses. $\lambda_r$ represents the rate at which a ligand-receptor bond cannot be formed, due to hits of ligand molecules with already busy receptors. By following a reasoning similar to the one reported above for evaluating $\lambda_i$, we can say that the state dependent rejection rate is equal to
\begin{equation}\label{Lambda_def_r}\lambda_{i,r} = Z^{*} c(t,r) i = \lambda_i \frac{n_b i }{\paren{R_{RX}-n_b} \paren{R_{RX}- i}}.\end{equation}
Clearly, the per state rejection rate is proportional to the number of busy receptor, $i$, and to the $Z^{*}$, which represents how frequently a collision between a particle and a busy receptor happens, which, in turn, is obviously proportional to $n_b$, and thus $Z^{*}= Z \frac{n_b}{R_{RX}-n_b}$. Consequently, the average rejection rate $\lambda_r$ can be evaluated as 
\begin{equation}\label{rejection}\lambda_r = \sum_{i=1}^{R_{RX}} {\lambda_{i,r} \pi_{i}}.\end{equation}

\subsection{Proposed congestion model}\label{mymodel}
We model the reception process of any single receptor as an M/M/1/1 queue (single server queueing system without waiting room, \cite{Kleinrock:1975:TVQ:1096491}). If we consider the quantities depicted in Fig. \ref{layout}, it is easy to express the distance from the center of TX to a generic point on the RX surface as a function of the angle $\alpha$ as
\begin{eqnarray}\label{distanza}
r(\alpha)=\sqrt{(d+R_{RX}sin(\alpha))^2+(R_{RX}cos(\alpha))^2},
\end{eqnarray}
\noindent with $\alpha \in [-\pi/2,\pi/2]$. 
As for the absorption rate of molecules to the node RX ($\lambda^{*}_a$), which is a measurable parameter, it is the superposition of the rates of arrival to each receptor $j$, denoted as $\lambda^{*}_{a,j}$. Each receptor $j$ is identified by the corresponding angle $\alpha_j$ and its distance from the TX center is equal to $r(\alpha_j)$. We denote as $j$=1 the receptor corresponding to $\alpha=\pi/2$, for which the distance to the center of TX is $d+r_{RX}$. Considering that, as in the previously illustrated model \cite{Pierobon2011sept}, the arrival rate is proportional to the local concentration of molecules, and taking into account (\ref{train}), it results that
\begin{eqnarray}\label{lambda_a}
\lambda^{*}_a=\sum_{j=1}^{R_{RX}} {\lambda^{*}_{a,j}}=\lambda^{*}_{a,1} (d+r_{RX}) \sum_{j=1}^{R_{RX}} {\frac{1}{r(\alpha_j)}},
\end{eqnarray}
where $\lambda^{*}_a$ is a system parameter which can be easily measured by the receiver. Thus, the assimilation rate for each receptor $\lambda^{*}_{a,j}$ can be derived by $\lambda^{*}_a$ by inverting (\ref{lambda_a}) and finding $\lambda^{*}_{a,1}$.

In an M/M/1/1 queue, given the rate of accepted traffic $\lambda_a$, the rate of the rejected traffic is $\lambda_r=\frac{T_{traff}\lambda_a^2}{1-T_{traff}\lambda_a}$. Thus, by using this result for each receptor, it is easy to find the system level rejection rate $\lambda^{*}_r$, given by
\begin{eqnarray}\label{lambda_r}
\lambda^{*}_r=\sum_{j=1}^{R_{RX}} {\lambda^{*}_{r,j}}={\lambda^{*}_{a,1}} (d+r_{RX}) \sum_{j=1}^{R_{RX}} {\frac{1}{r(\alpha_j)\left(\frac{r(\alpha_j)}{\lambda^{*}_{a,1} (d+r_{RX}) T_{traff}} - 1\right)}},
\end{eqnarray}
\noindent where $\lambda^{*}_{a,1}$ can be obtained by inversion of (\ref{lambda_a}), once the global arrival rate of molecules is known.

It is worth to note that although the system model illustrated in Fig. \ref{layout} assumes that the target of the drug is modeled as a spherical bio-nanomachine, in order to resemble some types of cells, the application of the proposed model is much more general. In more detail, the target of the drug can be modeled as a surface with a generic shape provided with receptors. If fact, as it appears from (\ref{lambda_a}) and (\ref{lambda_r}), the per-receptor assimilation and rejection rates are functions of the receptor distance from the center of transmission nanomachine, and can be easily evaluated once a measure $\lambda^{*}_a$ is available.  

If we assume that receptors are uniformly distributed on the RX surface, and by using the symmetry around the axis connecting the center of TX with the center of RX, it is possible to find more results. By partitioning the RX surface into $R_{RX}$ areas, it results that the surface surrounding each receptor has an average size $S=4 \pi r_{RX}^2/R_{RX}$. The angle $\alpha_1=\beta$ corresponding to the receptor located at distance $d+r_{RX}$ will be equal to $\beta=2/\sqrt{R_{RX}}$ due to geometrical considerations. 
If we divide the remaining surface of half semi-sphere (that for positive value of $\alpha$, the same reasoning holds for negative ones) in a number of spherical zones with angle $\Delta \alpha$, all receptors in these zones have the same distance from the center of TX, thus they see the same local concentration of drug molecules, and thus the same arrival rate $\lambda^{*}_{a,j}$. If we assume that the area $S$ dedicated to each receptor has roughly a square shape, it results that $S=(r_{RX} {\Delta \alpha})^2$, thus ${\Delta \alpha}=\sqrt{S}/r_{RX}=\sqrt{4 \pi/R_{RX}}$, and the number of spherical zones $F$ are given by $F=\frac{\pi-\beta}{2 {\Delta \alpha}}$. Since the area of each $k$-th spherical zone identified by $\alpha_k= k {\Delta \alpha}$ is given by $S_k=2 \pi r_{RX}^2 \left(sin(k {\Delta \alpha}) - sin((k-1) {\Delta \alpha}) \right)$ (see also Fig. \ref{layout}), the number of receptors in such a zone is 
\begin{eqnarray}\label{rece}
n_k=\frac{S_k}{S}=\frac{ R_{RX} \left(sin(k \sqrt{4 \pi/R_{RX}}) - sin((k-1) \sqrt{4 \pi/R_{RX}}) \right)}{2}
\end{eqnarray}
This means that (\ref{lambda_r}) becomes
\begin{eqnarray}\label{lambda_r_2}
\lambda^{*}_r={\lambda^{*}_{a,1}} (d+r_{RX}) \sum_{k=-F-1}^{F+1} {\frac{n_k}{r(\alpha_k)\left(\frac{r(\alpha_k)}{\lambda^{*}_{a,1} (d+r_{RX}) T_{traff}} -1\right)}},
\end{eqnarray}
\noindent and $\lambda^{*}_{a,1}$ can be obtained by
\begin{eqnarray}\label{lambda_a_1}
\lambda^{*}_{a,1}=\frac{\lambda^{*}_{a}}{ (d+r_{RX}) \sum_{k=-F-1}^{F+1} {\frac{n_k}{r(\alpha_k)}}}.
\end{eqnarray}
In order to consider a different distribution of receptors, by assuming that the symmetry with respect to $d$ still holds, it is enough to change the values of $n_k$, given that the condition $\sum_{k=-F-1}^{F+1} {n_k}=R_{RX}$ is satisfied.

\subsection{Application to drug delivery}\label{application}
In this sub-section we illustrate the application of our congestion model to drug delivery systems. Without loss of generality, we refer to the occupancy theory \cite{clark1933}, since it is the most commonly accepted theory. It says that the magnitude of drug response depends on the occupancy ratio of receptors by drug molecules. What we illustrate in what follows can similarly be adapted to the rate theory. 

As shown in \cite{Lambert01122004}, in some cases (e.g. competitive antagonist drugs) a full drug response can be produced even at low receptor occupancy, and it results that excess receptors, not bound to drug molecules, are no further needed to obtain the maximum drug response. Thus, in this case study we assume the usage of a drug type which produces the desired response when at least a fraction $f$ of receptors of the target cells are bound to the drug molecules. Thus, in order to suitably configure the drug delivery system, the first step consists of determining the minimum drug release rate that allows achieving the desired percentage of bound receptors.
By using the results shown in \cite{6967753}, it is easy to find that a good approximation of the arrival rate value $\lambda_o=\lambda^{*}_{a}+\lambda^{*}_{r}$ is given by
\begin{eqnarray}\label{arrival}
\lambda_o=\lambda^{*}_{a}+\lambda^{*}_{r}=\frac{r_{RX}}{d} \frac{R_{RX}r_{r,rx}}{\pi r_{RX} + R_{RX} r_{r,rx}} \frac{Q}{\Delta t}.
\end{eqnarray}

Thus, by applying (\ref{lambda_r_2}) and (\ref{lambda_a_1}) in (\ref{arrival}), it is easy to find a relationship between the release rate and $\lambda^{*}_{a}$. In addition, by using the Little's law \cite{Kleinrock:1975:TVQ:1096491}, it is immediate to find that the mean number of busy receptors (servers) in the system is given by $\lambda^{*}_{a} T_{traff}$, and the desired fraction $f$ results equal to the overall utilization coefficient $\rho$, which is equal to 
\begin{eqnarray}\label{ro}
\rho=\frac{\lambda^{*}_{a} T_{traff}}{R_{RX}}=f. 
\end{eqnarray}

\section{Performance Evaluation}\label{perf}
The performance of the system has been evaluated by using the BiNS2 simulator \cite{Felicetti2012,Felicetti2013172}, which is a Java package designed to simulate nano-scale biological communications in 3D. The approach of BiNS2 is fine grained: the position of each element, nanomachine or molecule, is evaluated at each simulation step, and collisions are managed according to a partially inelastic model, described by means of the coefficient of restitution \cite{Gidaspow2009}. Not only molecules, but also nanomachines can be either fixed or mobile. The surrounding environment can be unbounded or bounded. In the latter case, the bounding surface can have different shapes; currently the supported ones are sphere, cylinder, and cube, or a combination of them \cite{Felicetti201398}. In addition, BiNS2 allows modeling a receiver node with a finite number of receptors $R_{RX}$. Each receptor implements a finite, non-negligible reception time (trafficking time). When a molecule hits one of the molecule-compliant receptors, if it is not busy in another bond, it locks the relevant receptor for an amount of time whose distribution of the can be selected from a number of known statistics. In this work, the trafficking time is exponentially distributed with average value $T_{traff}$, whereas  the transmission rate at TX is fixed for the whole simulation duration. Instead, if the receptor is busy, the molecule is bounced back as it would have hit a portion of the surface of the receiver without receptors. 
In the scenario of this paper, nanomachines are fixed, whereas molecules move according to the Brownian motion and as results of collisions, which can occur with nanomachines or among themselves.

The simulation reliability of BiNS2 has been assessed by tuning simulation results with lab experiments \cite{Felicetti2013172}.
Finally, the BiNS2 package implements an octree-based computation approach \cite{Felicetti2013172}, which uses a dynamic splitting of the simulated environment into cubes of different size in order to parallelize the simulation, so as to benefit of the multi-thread capabilities of modern multi-core computer architectures, and thus strongly reducing the simulation time. In the simulation of the scenario considered in this paper, the simulation environment is unbounded; however, we implemented a cubic virtual boundary with side 1 mm to implement the octree algorithm. The TX nanomachine is located at the center of the cube. When molecules exit the cube, they are removed from the simulation, in order to limit the computational burden of the simulation. 
From the above description, it appears that BiNS2 implements exactly the receiver model that we consider in this paper. The main simulation parameters, together with their descriptions and values, are reported in Table \ref{tabella}. In the next subsection, we present the simulation results of the scenario described in section \ref{model}.  

\begin{table}[t]
\caption{Simulation parameters}
\centering
\scalebox{0.80}{
\begin{tabular}{l|l|l}
Symbol & Description & Value\\ \hline
$T$ & Temperature & 310 K \\
$e$ & Coefficient of restitution (part. inelastic collisions) & 0.95\\
$\eta$ & Viscosity & 0.0011 Pl \cite{Gidaspow2009}\\
$r_{RX}$ & Radius node RX & 2.5 $\mu m$ \\
$r_{TX}$ & Radius node TX & 2.5 $\mu m$ \\
$R_{RX}$ & Amount of surface receptors (node RX) & 10000 \\
$r_{c,tx}$ & Radius emitted molecules & 1.75 nm \\
$r_{r,rx}$ & Receptor radius (RX) & 4 nm \\
$T_{traff}$ & Trafficking time & 2 or 4 s \\
$\Delta t$ & Emission period (TX) & 20 ms \\
$d$ & RX-TX distance & 26.5 $\mu m$\\
\end{tabular}
\label{tabella}
}
\end{table}

\subsection{Numerical results}

\begin{figure}[!t]
\centering
\includegraphics[width=0.9\linewidth]{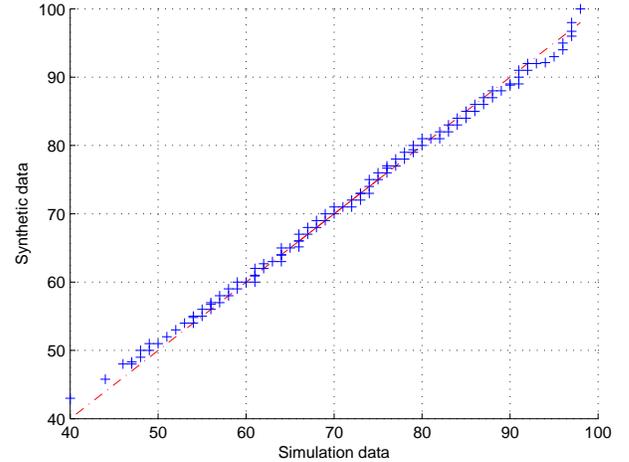}
 \caption{Q-Q plot of arrivals (both absorbed and rejected molecules in intervals of 0.1 s) obtained via simulation versus synthetic Poisson data with the same average.}
 \label{poiss_test}
\end{figure}

Before evaluating system performance, it is necessary to verify the suitability of the assumption of Poisson distributed arrivals. Fig. \ref{poiss_test} presents the Q-Q plot of simulation data versus synthetic Poisson data generated with the same average value. The simulation data include both absorbed and rejected molecules, gathered in intervals of 0.1 seconds. The resulting agreement is evident.

\begin{figure*}[!ht]
\centering
\includegraphics[width=1.0\linewidth]{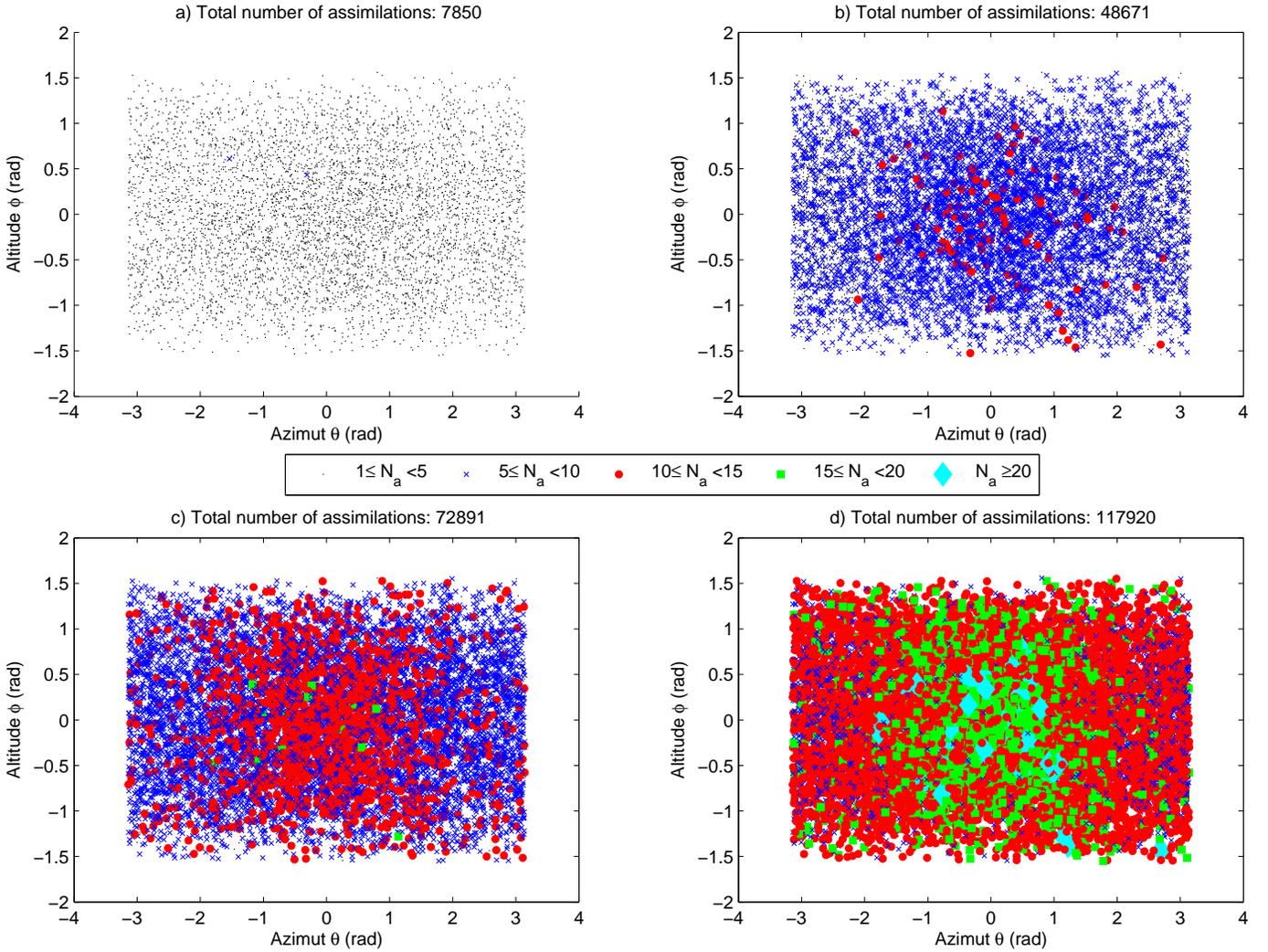}
 \caption{Maps of the number of assimilations for each surface receptor ($N_a$), as a function of its spherical coordinates, azimuth and the altitude, for increasing numbers of total assimilation: a) 7850, b) 48761, c) 72891, and d) 117920.}   
 \label{maps}
\end{figure*}

From the results in sections \ref{model}, it can be argued that the concentration of molecules is higher on the portion of the receiver surface facing the TX. In this portion of the surface, there are a lot of molecules which cannot establish a bond with receptors since most of them are already busy. Instead, the concentration of molecules close to the opposite portion of the receiver surface is much lower, since the molecules, following the negative gradient of concentration, tend to leave the RX node. As shown in section \ref{mymodel}, this phenomenon has a significant impact also on assimilations (and rejections). 
This observation is confirmed in Fig. \ref{maps}, which shows the map of assimilations for a trafficking time equal to $T_{taff}$=4s, at different times, corresponding to different amount of assimilations. $N_a$ represents the cumulative number of assimilations for each receptor, and it is plotted by means of a discrete value, chromatic scale for all the $R_{RX}$ receptors of RX. Each receptor is identified by its spherical coordinates, that is the azimuth $\theta$, whose range is (-$\pi$,$\pi$] radians, and the altitude $\phi$, whose range is (-$\pi/2$,$\pi/2$] radians. The direction connecting the center of TX and RX corresponds to $\paren{\theta,\phi}=\paren{0,0}$. As expected, the largest number of assimilations (i.e., $N_a$ values), in all considered samples, is located in the region around the center of the plot, that is $\paren{\theta,\phi}=\paren{0,0}$, which also represents the closest point between the RX and the TX. With respect to (\ref{conca}), the presence of the RX perturbs the concentration of molecules. In particular, those colliding with the portion of surface with values of $\theta$ and $\phi$ close to zero continue remaining close to the surface of the RX, due to negative gradient of the concentration, which ``pushes'' them towards large values of $d$. This facilitates the creation of ligand-receptor bonds on a larger area. However, in the region with coordinates close to $\paren{\theta,\phi}=\paren{0,0}$, the values of $N_a$ are often more than twice the value in the remaining of the RX surface. Since these receptors of the RX receive a larger portion of binding attempts, they are also responsible for most of rejections.
We now show simulation results obtained by using both the proposed model and the one illustrated in \cite{Pierobon2011sept}, suitably adapted to calculate the rejection rate, as described in section \ref{review}. In order to compare simulations with the theoretical models, we have assumed that they have the same mean arrival rate $\lambda_a=\lambda^{*}_{a}$ (and thus the same mean number of busy receptors). We have evaluated the rejection rate, since 
$\lambda_a$ is the only parameter that the RX can estimate reliably. In the simulation, we have tracked the number of rejection events, whereas the theoretical value  
in (\ref{rejection}) has been determined by evaluating the state probabilities numerically. As for the proposed model, the rate of rejection (i.e. unabsorbed drug molecules due to collisions with busy receptors) is evaluated by means of (\ref{lambda_r_2}). In addition, in the simulation we have verified that the ratio between the collisions of molecules with any of the deployed receptors, either busy or not, and those with the portion of surface of the RX not covered by receptors, closely match the theoretical value of  $R_{RX} \frac{\paren{r_{r,rx}+r_{c,rx}}^2}{4 r_{RX}^2}$, where $r_{r,rx}$ is the receptor radius, $r_{c,tx}$ is the molecules radius, and $r_{RX}$ is the RX node radius.

\begin{figure}[!ht]
\centering
\includegraphics[width=1.0\linewidth]{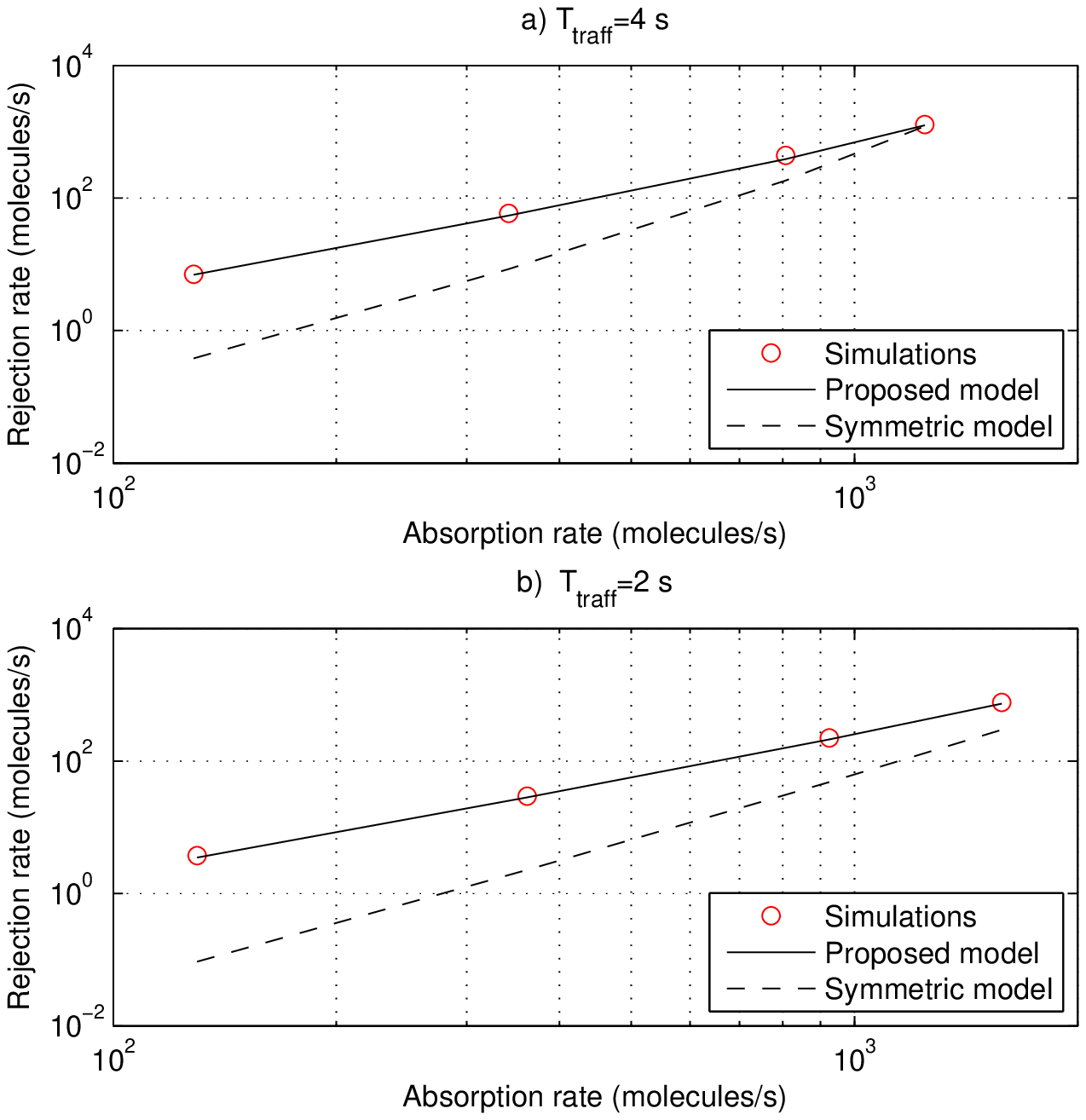}
 \caption{Rejections as a function of the absorption rate: a) $T_{traff}$ equal to  4 seconds, and b) $T_{traff}$ equal to 2 seconds.}
 \label{rejections}
\end{figure}

Results are shown in Fig. \ref{rejections}. In Fig. \ref{rejections}.a, the ordinate axis reports the rejection rate for the case $T_{traff}=4$ seconds, for both the simulations and the two theoretical models. The rejection rate derived from the model in \cite{Pierobon2011sept} and presented in section \ref{review} is labeled as ``Symmetric''.  Fig. \ref{rejections}.b  shows the same quantities for $T_{traff}=2$ seconds. In both figures, the abscissa reports the absorption rate measured at RX, expressed in drug molecules per second. Both abscissa and ordinate are expressed on a logarithmic scale. The first comment is that the symmetric model strongly underestimates the values of rejection rate $\lambda_r$, which means that it is not able to effectively detect congestion. Instead, our proposed model exhibits an excellent match with simulation results in all cases. This is an expected result, since in \cite{Pierobon2011sept} it is assumed that the concentration of molecules is the same all over the surface of the receiver. Instead, it could be not true, especially for low transmission rates and for short trafficking times. In addition, and this is the most important reason, receptors do not behave as a pool of servers, which can be invoked and used upon a new user enters the system (M/M/m/m model). Instead, they are isolated service elements, and this nature must be taken into account. Even if the symmetric model tries to mitigate this effect, by using a state dependent arrival process, it is not sufficient.
In particular, at low $T_{traff}$ values, the difference between the results of the symmetric model and the simulation increases, since it is expected that the RX is eager to assimilate drug molecules and to re-present available receptors. Nevertheless, this is true only in unlikely case of uniform concentration.
Instead, an interesting behavior appears in the case $T_{traff}$=4s when the absorption rate is very high, larger than 1000 molecules/s. In this case, as shown in Fig. \ref{rejections}.a, the estimation of the rejection rate given by the symmetric model tends to converge to the simulation values, which is also the value estimated by our model. Also this phenomenon can be easily explained. In fact, when the $T_{traff}$ value increases, the receiver is less prompt to free receptors. This means that when the absorption rate increases beyond a threshold value, due to a significant increase of the concentration of drug molecules nearby RX, even if the concentration close to the receiver surface is not homogeneous, it is so high that the limiting factor becomes the number of free receptors (about the 50\% in average), and thus the differences between the simulation and the two theoretical models tends to vanish. This trend is only barely visible for the case of $T_{traff}=2 s$, due to the fact that, at the maximum transmission rate, the average number of busy receptors is still low (about 32\%). In all the other cases, the difference between the symmetric and the proposed model is in the range of at least one order of magnitude.
\begin{figure}[!ht]
\centering
\includegraphics[width=1.0\linewidth]{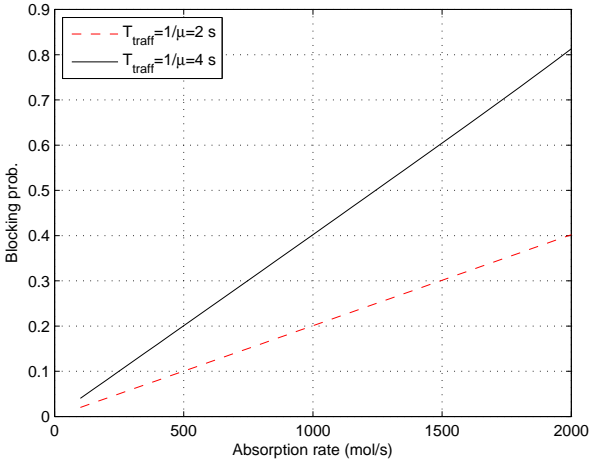}
 \caption{Blocking probability as a function of the absorption rate.}
 \label{l_a}
\end{figure}
\begin{figure}[!ht]
\centering
\includegraphics[width=1.0\linewidth]{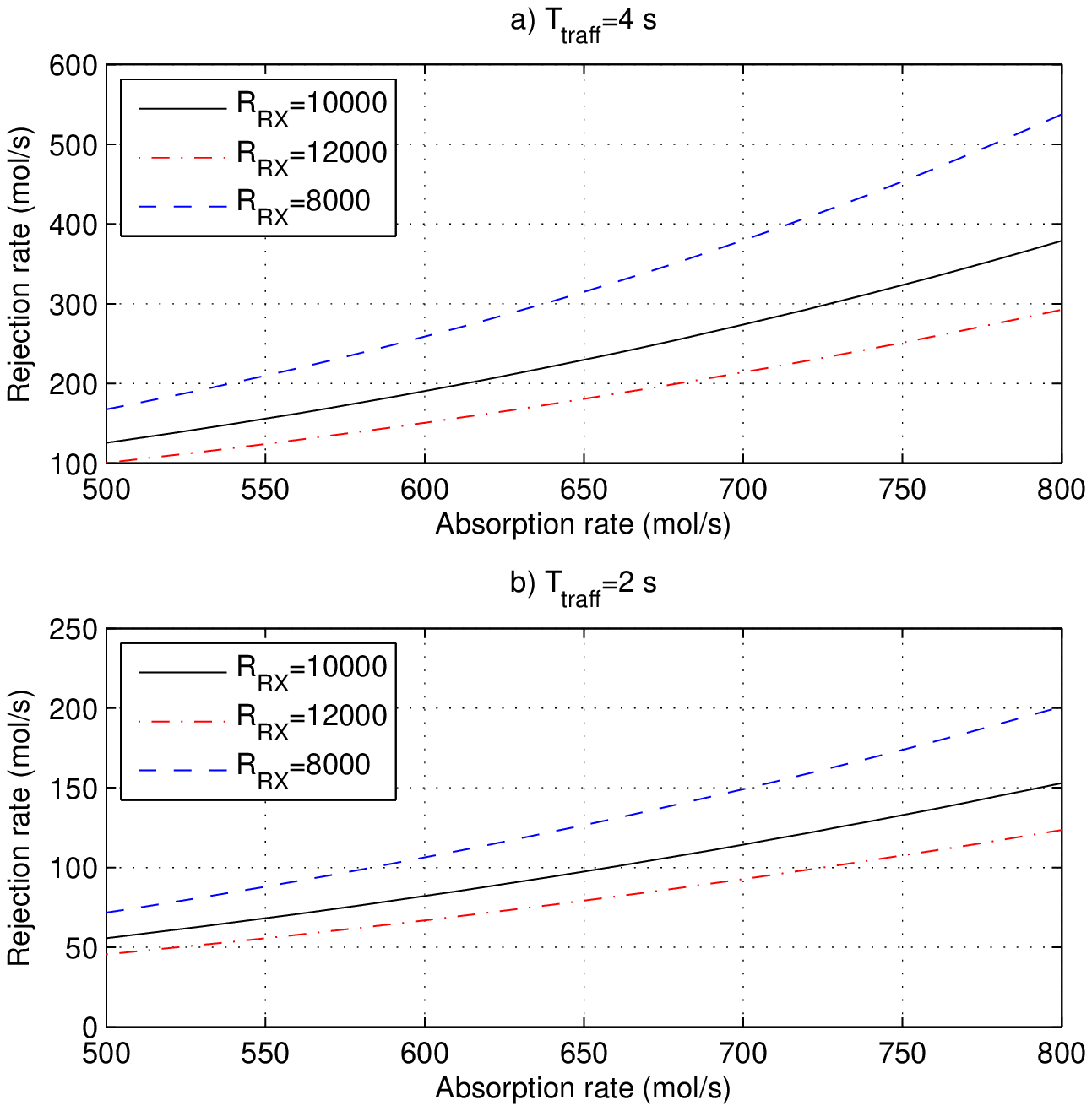}
 \caption{Rejection rate as a function of the absorption rate for different number of receptors and trafficking times.}
 \label{rece}
\end{figure}
Fig. \ref{l_a} shows the blocking probability, evaluated as $\frac{\lambda^{*}_r}{\lambda_o}$, where $\lambda_o=\lambda^{*}_r+\lambda^{*}_a$ is the system arrival rate. The abscissa reports the absorption rate, expressed in molecules per second. The net effect of the behaviors observed in the previous Fig. \ref{rejections} is a nearly liner increase of the blocking probability, as a function of the arrival rate, which is typical of pure loss systems in overload.
Finally, Fig. \ref{rece} shows the effect of varying the total number of receptors $R_{RX}$ by 20\% on the rejection rate, while keeping constant $\lambda^{*}_a$, indicated in the abscissa. As expected, when $R_{RX}$ increases, the rejection rate also decreases. Instead, when $R_{RX}$ decreases,  a marked increase of the rejection rate appears. Again, this phenomenon can be explained with the fact that receptors are not a multiplexed resource, but they are rather multiple, single resources. By increasing the distance between TX and RX, we have also investigated the occurrence of any marked decrease in the rejection rate, due to the fact that different concentration values should be less significant for large $d$ (see also Fig. \ref{maps}). However, we have found a nearly negligible decrease for the rate of rejected molecules with $d$ for $T_{traff}=4s$, not reported in this paper. This is probably due to the fact that the phenomenon visible in Fig. \ref{maps} is due to not only to the transmission distance, but also to the disturbing effect of the presence of the node RX itself. This is another confirmation that, when the system absorption rate $\lambda^{*}_a$ is kept constant, the factor with the most significant impact is the number of receptors, since it is not a multiplexed resource. 
\subsection{Lesson learned towards drug delivery}
In the previous subsection, we have analyzed the proposed model through simulations. Now, we highlight some key issues learned from this analysis, and focus on the operational procedure for designing an effective, localized drug delivery system. 

\begin{figure}[!ht]
\centering
\includegraphics[width=1.0\linewidth]{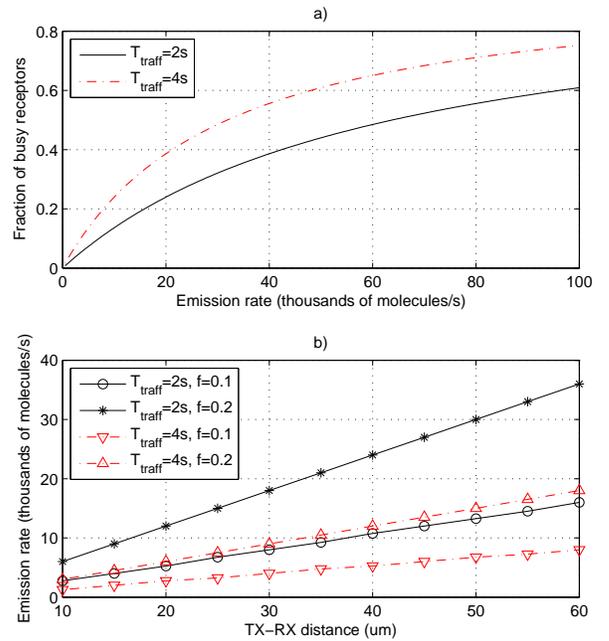}
 \caption{Percentage of busy receptors as a function of drug molecules release rate.}
 \label{balocco}
\end{figure}

First, in order to make use of the proposed model, it is necessary to characterize the underlying system. This can be done by means of lab experiments, or by using results already present in the literature. The goal of these experiments consists of estimating a small number of parameters, necessary for characterizing the system behavior. These parameters are the average trafficking time $T_{traff}$, and the number of surface receptors of the considered type, $R_{RX}$. Once the values of these two parameters are known, and the TX-RX distance $d$ is known, it is possible to make use of the model for any value of the emission rate $\frac{Q}{\Delta t}$ in order to evaluate the average number of busy receptors. Finally, according to the drug used, it is necessary to know the minimum fraction of receptors to be bound to drug molecules in order to maximize the drug effects. As mentioned above, for some types of drug this fraction could be quite low. For instance, in \cite{Lambert01122004}, the author shows that often only 5-10\% occupancy is needed to produce a full response when agonist drugs are used. For each drug, this value can be derived by in-vitro experiments. Thus, the final step consist of using the results obtained in subsection \ref{application} for estimating the optimal drug release rate $\frac{Q}{\Delta t}$, which guarantees the desired fraction of receptors bound to drug molecules, that is at least $f$, without drug overloading.
Fig. \ref{balocco}.a shows the percentage of busy receptors as a function of the release rate of drug molecules, for the trafficking time values already used in this paper, 2s and 4s, respectively. As expected, the value of $f$ increases with the rate $\frac{Q}{\Delta t}$. This sublinear increase is due to congestion. Fig. \ref{balocco}.b depicts the minimum release rate of drug molecules necessary to achieve the target occupancy $f$, versus the transmission range, for different values of the trafficking time. A first comment is that the dependency of drug release rate is nearly linear with the transmission range. In addition, the trafficking time has a significant influence on the slope of the release rate curve. In particular, the lower the trafficking time, the prompter is the target cell to absorb the drug, which makes it necessary to transmit more molecules to maintain the target occupancy $f$.


\section{Conclusion}\label{conclu}

In this paper, we have introduced the concept of congestion in diffusion-based molecular communications, and proposed a model for both illustrating the dynamical behavior of the phenomenon and analyzing the root causes of it. The considered continuous emission of molecules is typical of localized drug delivery systems.  Our proposed model of the receiver nanomachine includes an M/M/1/1 queue for each receptor. This model can capture the dependency of the arrival rate (ligand-receptor bond formation) on both transmission range and nature of receptor. Receptors cannot be simply considered a set of multiplexed resources such as a server pool. They behave as isolated, multiple nano-receivers, independent of each other. An extensive numerical analysis shows that this model is effective in modeling congestion conditions, especially in the range of distances typical of diffusion-based molecular communication.

The results of this work can be used for implementing rate control algorithms for molecular communications, in particular for drug delivery systems. We have detailed the operational procedure for determining the suitable drug delivery rate from a set of implanted emitting nanomachines, which release drug nanoparticles close to the target cells. 


In addition, since the arrival rate can be measured by the receiver bio-nanomachine by counting the absorbed molecules, 
the system can also be designed with adaptive features. In particular, it is possible to add a further controller nanomachine, which can estimate whether the used transmission rate is optimal, and adapt its value to any change of the operational conditions. Release rate adjustments can be triggered by using feedback messages \cite{nakano13} sent when the observed deviations from the original design conditions are significant.

The drug delivery system analyzed in this paper is analyzed in a quasi-static environment, such as the extracellular matrix. Drug delivery in more dynamic environments, such as blood vessels, we will be the object of future investigations. In such scenarios different configurations can be studied, such as fixed transmitters, anchored to vessel wall, and mobile targets (e.g. circulating tumor cells), or mobile transmitters and mobile targets, or mobile transmitters and fixed targets (e.g. solid tumors). 

%
\bibliographystyle{IEEEtran}
\bibliography{Femminella_fin}  
\end{document}